\documentclass[pra,superscriptaddress,twocolumn,longbibliography
]{revtex4-1}

\usepackage{amsmath,amsfonts,amssymb}
\usepackage{dsfont}
\usepackage{graphicx,bm}
\usepackage{txfonts}
\usepackage{color}
\usepackage{hyperref}

\usepackage{pict2e,picture,graphicx}

\makeatletter
\DeclareRobustCommand{\Arrow}[1][]{%
\check@mathfonts
\if\relax\detokenize{#1}\relax
\settowidth{\dimen@}{$\m@th\rightarrow$}%
\else
\setlength{\dimen@}{#1}%
\fi
\sbox\z@{\usefont{U}{lasy}{m}{n}\symbol{41}}%
\begin{picture}(\dimen@,\ht\z@)
\roundcap
\put(\dimexpr\dimen@-.7\wd\z@,0){\usebox\z@}
\put(0,\fontdimen22\textfont2){\line(1,0){\dimen@}}
\end{picture}%
}
\makeatother
\newcommand{\veryshortrightarrow}{\hspace{.2mm}\scalebox{.8}{\Arrow[.1cm]}\hspace{.2mm}}

\newcommand{\abs}[1]{\left|{#1}\right|}

\newcommand{\pt}[1]{\left( #1 \right)}
\newcommand{\pq}[1]{\left[ #1 \right]}
\newcommand{\pg}[1]{\left\{ #1 \right\}}

\newcommand{\ee}{{\rm e}}
\newcommand{\ii}{{\rm i}}

\usepackage{bbold}

\definecolor{green}{rgb}{0,0.6,0}

\usepackage[normalem]{ulem}

\newcommand{\stkout}[1]{\ifmmode\textrm{\sout{\ensuremath{#1}}}\else\sout{#1}\fi}

\DeclareMathOperator\erfc{erfc}

\begin{document}

\title{Microwave quantum illumination with correlation-to-displacement conversion}

\author{Jacopo Angeletti}
\affiliation{Physics Division, School of Science and Technology, University of Camerino, I-62032 Camerino (MC), Italy}
\affiliation{Department of Physics, University of Naples ``Federico II", I-80126 Napoli, Italy}
\affiliation{INFN, Sezione di Perugia, I-06123 Perugia, Italy}
\affiliation{Ming Hsieh Department of Electrical and Computer Engineering, University of Southern California, Los
Angeles, California 90089, USA}

\author{Haowei Shi}
\affiliation{Ming Hsieh Department of Electrical and Computer Engineering, University of Southern California, Los
Angeles, California 90089, USA}

\author{Theerthagiri Lakshmanan}
\affiliation{Physics Division, School of Science and Technology, University of Camerino, I-62032 Camerino (MC), Italy}
\affiliation{Department of Physics, University of Naples ``Federico II", I-80126 Napoli, Italy}

\author{David Vitali}
\affiliation{Physics Division, School of Science and Technology, University of Camerino, I-62032 Camerino (MC), Italy}
\affiliation{INFN, Sezione di Perugia, I-06123 Perugia, Italy}
\affiliation{CNR-INO, I-50125 Firenze, Italy}

\author{Quntao Zhuang}
\email{qzhuang@usc.edu}
\affiliation{Ming Hsieh Department of Electrical and Computer Engineering, University of Southern California, Los
Angeles, California 90089, USA}
\affiliation{Department of Physics and Astronomy, University of Southern California, Los Angeles, California 90089, USA}

\begin{abstract}
Entanglement is vulnerable to degradation in a noisy sensing scenario, but surprisingly, the quantum illumination protocol has demonstrated that its advantage can survive. However, designing a measurement system that realizes this advantage is challenging since the information is hidden in the weak correlation embedded in the noise at the receiver side. Recent progress in a correlation-to-displacement conversion module provides a route towards an optimal protocol for practical microwave quantum illumination. In this work, we extend the conversion module to accommodate experimental imperfections that are ubiquitous in microwave systems. To mitigate loss, we propose amplification of the return signals. In the case of ideal amplification, the entire six-decibel error-exponent advantage in target detection error can be maintained. However, in the case of noisy amplification, this advantage is reduced to three-decibel. We analyze the quantum advantage under different scenarios with a Kennedy receiver in the final measurement. In the ideal case, the performance still achieves the optimal one over a fairly large range with only on-off detection. Empowered by photon number resolving detectors, the performance is further improved and also analyzed in terms of receiver operating characteristic curves. Our findings pave the way for the development of practical microwave quantum illumination systems.
\end{abstract}

\date{\today}

\maketitle

\section{Introduction}
Quantum illumination (QI) is an entanglement-assisted sensing scheme that enhances the precision and sensitivity of target detection 
~\cite{lloyd2008enhanced,tan2008quantum,shapiro2020quantum}, via entangling the signal probes with locally stored idlers. Originally developed to simply detect the presence or absence of a target, QI offers a $6$-decibel improvement in error exponent due to entanglement~\cite{tan2008quantum}. In recent years, QI has been extended to improve target range and angle detection~\cite{zhuang2021quantum,zhuang2022ultimate}, demonstrating an even greater advantage over classical counterparts in the intermediate signal-to-noise-ratio (SNR) region, thanks to the threshold phenomena of nonlinear parameter estimation~\cite{zhuang2022ultimate}.

Despite these theoretical advancements in QI, its experimental realization in the microwave domain, which is the natural scenario for its application, has faced several limitations. One of the practical challenges is the need for extensive cooling for microwave quantum-limited detection, due to the high natural noise background, and the lack of developed photon-counting detection technology~\cite{dixit2021,assouly2022demonstration}. To address these issues, a solution for QI based on optical-microwave transduction has been proposed~\cite{Barzanjeh_2015}. This approach utilizes an optical idler mode for noiseless storage at room temperature, and up-converts the microwave return mode to the optical domain for quantum-limited joint detection of optical photons. However, the current state-of-the-art efficiency in optical-microwave transduction~\cite{lauk2020perspectives,awschalom2021development,fan2018superconducting,han2021microwave,Brubaker2022omc,sahu2023entangling} falls short of what is required to sustain this transduction-based scheme in the near future.

In addition to the practical challenges, a fundamental limitation of QI is the receiver design problem. Currently, practical receivers such as the optical parametric amplifier receiver (OPAR) and the phase-conjugate receiver (PCR) can only attain half of the error exponent advantage~\cite{Guha2009}. The optimal receiver would require unit-efficiency sum-frequency-generation at the single photon level~\cite{zhuang2017optimum}, which is highly challenging to realize experimentally. The problem of optimal receiver design seems to necessitate nonlinear processes and joint operations on the idler and return modes, making it difficult to implement in practice.

Previous in-principle demonstrations of QI target detection have been hindered by the aforementioned limitations. One example is an optical domain simulation, which injected noise to mimic a microwave scenario and utilized a sub-optimal OPAR~\cite{zhang2015}. This approach achieved approximately $20\%$ of the error exponent advantage. Another demonstration in the microwave domain used a digitally reconstructed PCR~\cite{Barzanjeh_2020}, but was unable to surpass the performance of the classical benchmark represented by an ideal coherent state source with the same mean number of photons and homodyne detection. More recently, the OPAR scheme was adapted to the microwave domain, overcoming several challenges in microwave photon processing~\cite{assouly2022demonstration} and again yielding roughly $20\%$ of the error exponent advantage.

A recent development in the field of optimal receiver design is the correlation-to-displacement (`$\rm C\veryshortrightarrow D$') conversion proposal, which suggests that the optimal receiver design can be achieved by heterodyne-detecting the return mode separately and processing the associated conditional idler field~\cite{shi2022fulfilling}. Upon heterodyne detection of the return modes, the idler modes collapse to coherent states embedded in weak thermal noise. With the help of well-established coherent state discrimination protocols, the C$\veryshortrightarrow $D receiver design can attain the optimal error probability of QI~\cite{Nair:20}. This receiver design requires only programmable linear optics~\cite{Ma2011tunablebeamspl,Kraftmakher2020tunablemwsplit} and photon detection, making it more feasible for experimental realization. Additionally, it eliminates the need for mode-matching between the noisy return fields at room temperature and the cooled idler fields, avoiding technical difficulties.
\begin{figure*}
\centering
\includegraphics[width=0.75\linewidth]{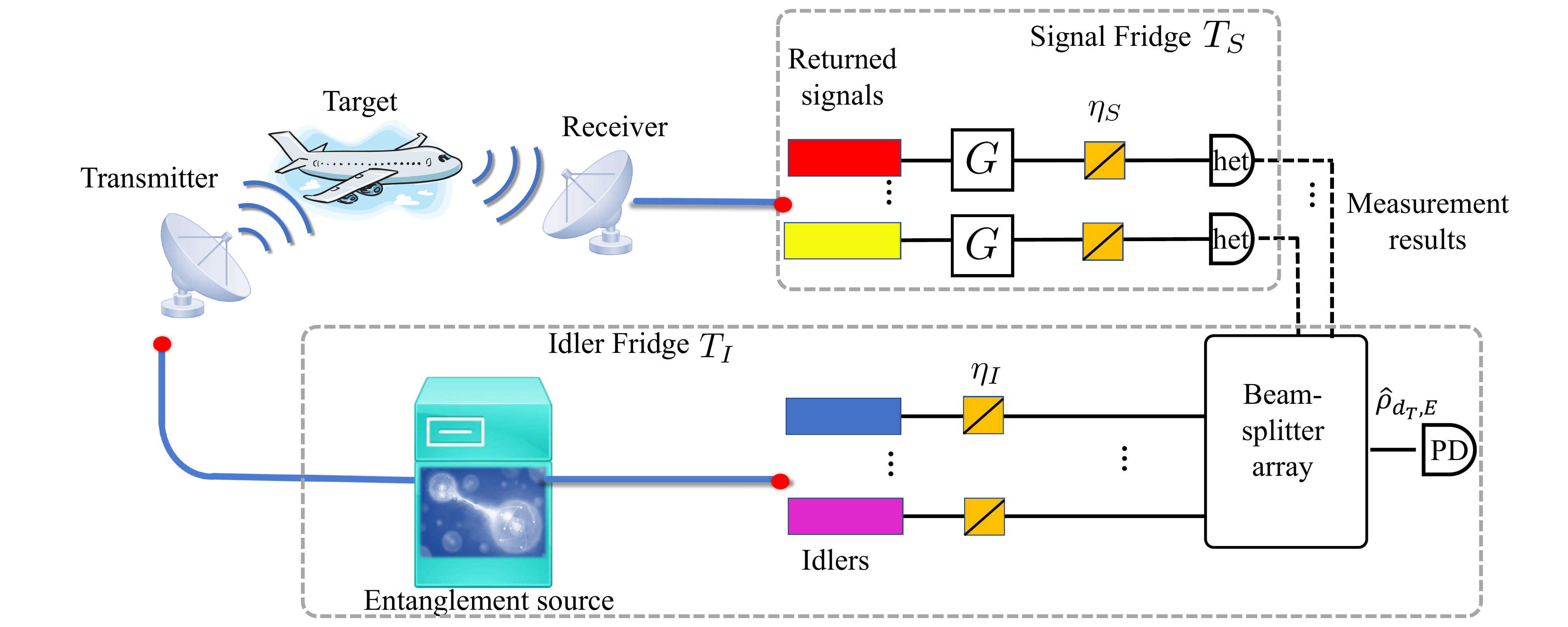}
\caption{Schematic of the quantum illumination, with a practical receiver based on correlation-to-displacement conversion, in presence of noise and loss. `het': heterodyne detection. `PD': photo-detection.
}
\label{cartoon}
\end{figure*}
In this study, we evaluate the feasibility of the C$\veryshortrightarrow$D receiver design in the microwave domain. We account the lossy antenna coupling to the detection in real radar systems, by introducing loss $1-\eta_S\le 1$ in the return mode prior to heterodyne detection. To mitigate this loss, we suggest using parametric amplification with gain $G \geq 1$. Our results show that the full optimal six-decibel error-exponent advantage can be retained when $G \eta_S\gg1$ if the amplifier is quantum limited. Even if the amplifier introduces noise at room temperature, the C$\veryshortrightarrow$D receiver still provides a three-decibel advantage over the ideal classical system. Furthermore, we consider the case of limited detection capability in the idler modes. Instead of the complex Dolinar receiver, we consider the simpler Kennedy receiver and still observe the optimal error exponent advantage. Finally, we compare the practical C$\veryshortrightarrow$D receiver design with both the classical coherent-state homodyne detection and the PCR (which is more effective than the OPAR~\cite{shi2020practical}).

This paper is organized as follows. Sec.~\ref{ProtSec} describes the protocol, while Sec.~\ref{QITDSec} recalls the basic properties and tools of QI. Sec.~\ref{Sec_VIAHet} provides a brief review of the C$\veryshortrightarrow$D receiver and its performance under ideal conditions. Sec.~\ref{Practmw_sec} discusses relevant experimental limitations in the case of microwave QI, and Sec.~\ref{nic-sec} shows the performance of the C$\veryshortrightarrow$D module in the presence of such realistic scenarios. Sec.~\ref{Perf-sec} compares the performance of the C$\veryshortrightarrow$D module with that of classical QI based on coherent state and homodyne detection and that of the PCR. In Sec.~\ref{sec:number_resolving}, we consider performance enhancement if we further allow number-resolving detection. In particular, Sec.~\ref{roc-sec} presents the Neyman-Pearson framework and receiver operating characteristic (ROC) curves. Finally, we conclude the paper in Sec.~\ref{concl}.

\section{Overall protocol}\label{ProtSec}
As shown in Fig.~\ref{cartoon}, in a target detection scenario, the transmitter sends signals to the target, and then the receiver collects return signals and performs measurement to infer about target's presence or absence. To benefit from entanglement, a source generates pairs of idler-signal entangled pulses. The idlers are stored locally and used to assist joint measurements with the return signals. In QI, such signal-idler entanglement provides a six-decibel error exponent advantage, despite being destroyed by extremely lossy transmission and high noise background.

Our proposed receiver system adapts the $\rm C\veryshortrightarrow D$ conversion approach to practical receiver operating conditions. While the idlers are cooled to $T_I\sim 10$ mK to enable quantum advantage, the returned signal part is cooled to a much higher temperature $T_S$ for experimental convenience. Such a layout is possible as the $\rm C\veryshortrightarrow D$ conversion module only feeds the classical heterodyne measurement results on the `warm' and noisy returned signals, to perform conditional linear optical transforms on the `cool' idler alone (indicated by the dashed line), avoiding idler contamination. Finally, photo-detection is performed on the transformed idler, and a decision on the target's presence or absence is made according to the measurement result. To compensate for additional loss $1-\eta_S$ at the receiver antenna, amplification of gain $G$ is performed. However, the loss $1-\eta_I$ on the idler needs to be minimized and cannot be compensated. The photo-detection can be realized via coupling the microwave idler modes to transmon qubits, as demonstrated in Refs.~\cite{assouly2022demonstration,dixit2021}.

\section{Quantum illumination for target detection}\label{QITDSec}
QI is a quantum-based remote sensing technique that leverages the entanglement between signal $\pt{a_S}$ and idler $\pt{a_I}$ modes. The signal mode probes a target region, while the idler one is kept at the emission station. By performing a joint measurement on the signal and idler modes, the quantum correlations of the transmitted state are exploited at the receiving station. The problem is framed as a binary decision-making task, where the two hypotheses are: `target absent' $\pt{H_0}$ and `target present' $\pt{H_1}$. The asymptotic optimal input state is a two-mode squeezed vacuum (TMSV) state, a bipartite Gaussian state characterized by its covariance matrix (CM)~\cite{Nair:20,DePalma2018minimumerror}
\begin{equation}\label{VSI}
    \begin{split}
    	\textbf{V}_{SI}&=
    		\left(
    		\begin{array}{cc}
    			\pt{2N_S+1}\textbf{I}	 &2\sqrt{N_S\left(N_S+1\right)}\textbf{Z} 	\\
    			2\sqrt{N_S\left(N_S+1\right)}\textbf{Z} &2\pt{N_S+1}\textbf{I} 	
    		\end{array}
    		\right),
  \end{split}
\end{equation}%
where $\textbf{Z} = {\rm diag}\{1,\,-1\}$, $\textbf{I} = {\rm diag}\{1,\,1\}$, and $\left\langle a_S^\dag{}a_S\right\rangle=N_S$ is the signal brightness.
While the idler is stored for later detection, the signal is transmitted through a phase-shift thermal-loss channel $\Phi_{\kappa,\,\theta}$, whose action on its mode when the target is present is described by
\begin{equation}\label{ar}
    a_R=\ee^{\ii\theta}\sqrt{\kappa}a_S+\sqrt{1-\kappa}a_B,
\end{equation}%
while the absence of a target corresponds to the case $\kappa = 0$, i.e., where the channel is $\Phi_{0,\,0}$. Upon the channel $\Phi_{\kappa,\,\theta}$, the CM Eq.~(\ref{VSI}) becomes
\begin{equation}\label{VRI}
    \begin{split}
    	\textbf{V}_{RI}&=
    		\left(
    		\begin{array}{cc}
    			\pq{2\pt{\kappa N_S+N_B}+1}\textbf{I}	 &2\sqrt{\kappa N_S\left(N_S+1\right)}\textbf{R}\textbf{Z}\\
    			2\sqrt{\kappa N_S\left(N_S+1\right)}\textbf{Z}\textbf{R}^T &\pt{2N_S+1}\textbf{I} 	
    		\end{array}
    		\right),
    \end{split}
\end{equation}%
where $\textbf{R}\textbf{Z}=\Re\pq{\ee^{\ii\theta}\pt{\textbf{Z}+\ii\textbf{X}}}$ (with $\Re$ indicating the real part and $\textbf{X}$ the Pauli-X matrix), such that $\textbf R$ denotes a phase rotation of $-\theta$,
and $\left\langle a_B^\dag{}a_B\right\rangle=N_B/\pt{1-k}$ is the mean number of thermal background photons. Tab.~\ref{tab1} shows the mean thermal photon number for a typical microwave field at $\omega=2\pi\times 5$ GHz at temperatures of interest. The signal and return modes propagate at room temperature, while---depending upon the chosen device---detectors and amplifiers can be operated at temperature $T_S$ equaling either the room temperature, a few Kelvins, or ideally close to the Josephson parametric amplifier generating the TMSV state at microwave frequency~\cite{Flurin2012,Abdo2013}, which is typically placed in the cold plate of a dilution refrigerator at about $10$ mK~\cite{Barzanjeh_2020,assouly2022demonstration,Sandbo2019QENR}. The idler is always stored in the dilution refrigerator at about $T_I\sim 10$ mK~\cite{Barzanjeh_2020,assouly2022demonstration,Sandbo2019QENR}, to enable quantum advantages.
\begin{table}[t!]
    \centering
    \begin{tabular}{ccc}
    \hline
        $\omega/2\pi\,\pq{\textrm{GHz}}$ &  $T\,\pq{\textrm{K}}$ & $N\sim$\\
        \hline
        $5$ & $3\times 10^2$ & $1.25\times10^3$\\
         & $10^2$ & $4.15\times 10^2$\\
         & $10$ & $40$\\
         & $4$ & $15$\\
         & $1$ & $4$\\
         & $10^{-1}$ & $10^{-1}$\\
         & $10^{-2}$ & $4\times 10^{-11}$\\
         & $4\times10^{-3}$ & $9\times 10^{-27}$\\
    \end{tabular}
    \caption{Values of mean thermal photon numbers for a microwave mode at $\omega=2\pi\times 5$ GHz at temperature values of interest.}
    \label{tab1}
\end{table}

\section{Correlation-to-displacement conversion in the ideal case}\label{Sec_VIAHet}
Ref.~\cite{shi2022fulfilling} proposes a conversion module for capturing and transforming quantum correlation into coherent quadrature displacement, to enable the optimal receiver design for various entanglement-enhanced protocols. The module is based on heterodyne and programmable passive linear optics, and maps the multi-mode quantum detection problem to the semi-classical detection problem of a single-mode noisy coherent state, allowing for explicit measurements to achieve the optimal performance. The module provides a paradigm for processing noisy quantum correlations for near-term implementation and can be applied to a wide range of entanglement-enhanced protocols, including quantum illumination, phase estimation, classical communication, target ranging, and thermal-loss channel pattern classification.

QI for target detection considers the discrimination between two channels, $\Phi_{0,\,0}$ and $\Phi_{\kappa,\,0}$. In the ideal case, the conversion module produces the displaced thermal states $\rho_{0,\,N_S}$ (target absent, $H_0$) and $\rho_{\sqrt{x},\,E}$ (target present, $H_1$), where $x\sim P^{\pt{M}}\pt{\,\cdot\,;\,\xi_{\rm Ideal}}$ obeys a (generalized) $\chi^2$ distribution with  $\xi_{\rm Ideal} \equiv \kappa N_S(N_S+1)/2 (\kappa N_S+N_B+1)$. Here the probability density function for the $\chi^2$ distribution parameterized by $\xi$ is given by
\begin{equation}\label{PkMx}
    P^{\pt{M}}\pt{x;\,\xi}=\frac{x^{M-1}\ee^{-x/\pt{2\xi}}}{\pt{2\xi}^{M}\Gamma\pt{M}},    
\end{equation}%
where $\Gamma\pt{M}=\pt{M-1}!$ is the gamma function. This leads to the error probability performance limit
\begin{equation}\label{PCD}
    P_{\rm C\veryshortrightarrow D}=\int_0^{+\infty} dx\,P^{\pt{M}}\pt{x;\,\xi}P_\textrm{H}\pt{\rho_{0,\,N_S},\,\rho_{\sqrt{x},\,E}},
\end{equation}%
where $P_\textrm{H}$ is the Helstrom limit~\cite{Helstrom_1967, Helstrom1969, Helstrom_1976} 
\begin{equation}
    P_\textrm{H}\pt{\rho_1,\rho_2}=\frac{1}{2}\pt{1-\frac{1}{2}\textrm{Tr}\pq{\abs{\rho_1-\rho_2}}},
\end{equation}
in the case of equal prior probability. As shown in Ref.~\cite{shi2022fulfilling}, even though the exact solution of Eq.~\eqref{PCD} is challenging, we can obtain lower (LB) and upper bounds (UB) for the error exponent $r_{\rm C\veryshortrightarrow D}= -\lim_{M\to \infty}\ln\pt{ P_{\rm C\veryshortrightarrow D}}/M$. The upper bound can be achieved by approximating $\rho_{\sqrt{x},\,E}$ as a coherent state and $\rho_{0,\,N_S}$ as vacuum. In the respect of the asymptotic analysis, the Helstrom limit approaches $P_\textrm{H}\pt{\rho_{0,\,N_S},\,\rho_{\sqrt{x},\,E}}\sim\ee^{-x}/4$, which---combined with Eq.~(\ref{PCD})---gives the upper bound $r^{\pt{\textrm{UB}}}_{\rm C\veryshortrightarrow D}=2\xi$. On the other hand, a lower bound of the conversion module performance can also be obtained as~\cite{shi2022fulfilling}
\begin{equation}\label{rCDeq}
    r_{\rm C\veryshortrightarrow D}^{(\rm LB)}=2\xi\pt{\sqrt{N_S+1}-\sqrt{N_S}}^2.
\end{equation}
In comparison, the optimal classical case, achieved when a coherent-state with mean photon number $N_S$ is sent to the target, has the error exponent
\begin{equation}\label{rCSeq}
    r_\textrm{CS}=\kappa N_S\pt{\sqrt{N_B+1}-\sqrt{N_B}}^2.
\end{equation}
In the $N_S\ll1$ and $N_B\gg1$ limit, one finds that $r^{\pt{\textrm{UB}}}_{\rm C\veryshortrightarrow D}\simeq r^{\pt{\textrm{LB}}}_{\rm C\veryshortrightarrow D}\simeq 4r_\textrm{CS}$, which achieves the optimal advantage.
\begin{figure}[t!]
\centering
\includegraphics[width=.9\linewidth]{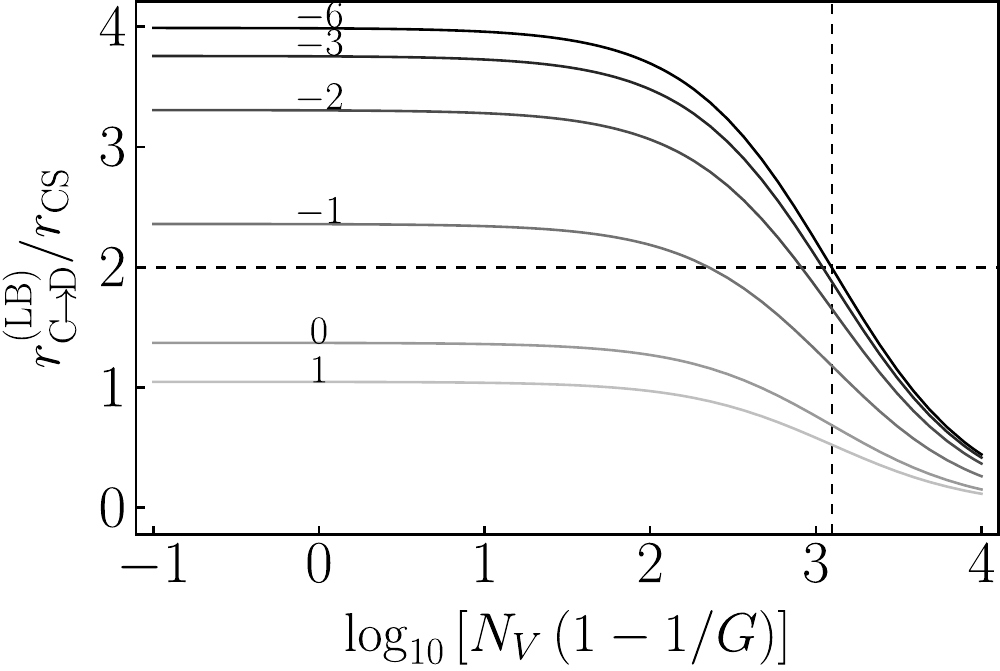}
\caption{Behavior of $r_{\rm C\veryshortrightarrow D}^{\pt{\textrm{LB}}}/r_{\textrm{CS}}$ as a function of $\log_{10}\pq{N_V\pt{1-1/G}}$ with amplification and ideal signal and idler detection, for different values of $\log_{10}N_S$, given $N_B=1250$ and $\kappa=0.01$. $N_S=\pg{10^{1},\,10^0,\,10^{-1}, 10^{-2},\,10^{-3},\,10^{-6}}$ from bottom to top, as indicated by the labels on top of the curves. The plot indicates that the amplification stage provides a factor of advantage greater than $2$ (as indicated by the horizontal dashed line) for a range of relevant parameters. This is due to the robust compensation of noise effects achieved by amplifying, as demonstrated by the vertical dashed line at $N_V\pt{1-1/G}=N_B$.
}
\label{Fig2}
\end{figure}

\section{Practical microwave detection scenario} \label{Practmw_sec}
Regardless of the technology or setup employed, non-idealities or imperfections will always exist in practical systems, affecting their performance. To mitigate this, we propose the use of a pre-detection amplifier, which can compensate for additional coupling loss. Our results demonstrate that this approach can effectively improve the performance of binary hypothesis testing and enhance the accuracy of state discrimination.

Before detection, the returned mode is amplified using a quantum amplifier, leading to the amplified mode
\begin{equation}\label{aa}
    a_A=\sqrt{G}a_R+\sqrt{G-1}a_V^\dag{},
\end{equation}%
where $\left\langle a_V^\dag{}a_V\right\rangle=N_V$ is the mean photon number of the amplifier noise mode. The amplified $a_A$ and the idler $a_I$ modes share the CM
\begin{equation}\label{VAImat}
	\textbf{V}_{AI}=
		\left(
		\begin{array}{cc}
			\pt{2N_A+1}\textbf{I}	 &V_{12}\textbf{R}\textbf{Z}\\
			V_{12}\textbf{Z}\textbf{R}^T &\pt{2N_S+1}\textbf{I} 	
		\end{array}
		\right),
\end{equation}%
where
\begin{equation}\label{V12}
    \begin{split}
        N_A=&\,\left\langle a_A^\dag{}a_A\right\rangle\\
        =&\,G\pq{\kappa N_S+N_B+\pt{1-1/G}\pt{N_V+1}},\\
        V_{12}=&\,2\sqrt{G\kappa N_S\left(N_S+1\right)}.
    \end{split}
\end{equation}%
Microwave amplifiers with gain $G\sim 100$ and excess noise of $N_V\sim 10$ photons have been successfully utilized in various microwave QI experiments~\cite{Barzanjeh_2020}. Additionally, superconducting quantum computers employ microwave quantum-limited amplifiers that exhibit added noise levels of about half a photon~\cite{IBMweb}. The behavior of such experimental systems can be accurately described by the phase-insensitive linear amplifier model presented in Eq.~\eqref{aa}.

It should be noted how, comparing Eq.~\eqref{VAImat} with the one without any amplification Eq.~\eqref{VRI}, the performance lower bound Eq.~\eqref{rCDeq} applies also to the case with the amplifier, as long as one replaces the parameters $\kappa\to G\kappa$ and $N_B\to N_A-G\kappa N_S$. Furthermore, we see that if $\pt{1-1/G}\pt{N_V+1}\ll N_B$, the performance of the conversion module does not change asymptotically. This is verified in Fig.~\ref{Fig2} via calculating $r_{\rm C\veryshortrightarrow D}^{\pt{\textrm{LB}}}/r_\textrm{CS}$ vs $\log_{10}\pq{N_V\pt{1-1/G}}$, where the factor of four ($6$ dB) advantage is seen at the $N_S\ll1$ limit.

The same analysis can also be applied to the non-ideal scenario of imperfect heterodyne detection of the amplified mode and imperfect idler detection. Heterodyne detection efficiency in the microwave regime typically ranges from $40\%$ to $70\%$ depending on the input power. 
However, in the scope of our analysis, $\eta_S$ represents the overall channel efficiency, which is dependent on the specific experiment and may be much lower, with realistic values around $10\%$ or even less (down to $1\%$).


For simplicity, we assume the non-ideal heterodyne detection to be symmetric in the quadratures, resulting in the input-output relation
\begin{equation}\label{apa}
    a'_A=\sqrt{\eta_S}a_A+\sqrt{1-\eta_S}a_{E_1},
\end{equation}%
where we set $\left\langle a_{E_1}^\dag{}a_{E_1}\right\rangle=N_{E_1}$. By performing the analysis through channel composition [see Eqs.~\eqref{ar},~\eqref{aa}, and~\eqref{apa}], one can obtain
\begin{equation}
    \begin{split}
        a'_A&=\ee^{\ii\theta}\sqrt{\eta_SG\kappa}a_S+\sqrt{1-\eta_SG\kappa}\tilde a,\\
        \tilde a&=\frac{\sqrt{\eta_SG\pt{1-\kappa}}a_B+\sqrt{\eta_S\pt{G-1}}a_V^\dag{}+\sqrt{1-\eta_S}a_{E_1}}{\sqrt{1-\eta_SG\kappa}},
    \end{split}
\end{equation}%
with $\pq{\tilde a,\tilde a^\dag{}}=1$, and
\begin{equation}
        \left\langle\tilde a^\dag{}\tilde a\right\rangle=\frac{\eta_SGN_B+\eta_S\left(G-1\right)\left(N_V+1\right)+\pt{1-\eta_S}N_{E_1}}{1-\eta_SG\kappa}.
\end{equation}%
With this composition, the channel is now characterised by the parameters
\begin{equation}\label{chcompsub}
    \begin{split}
        \kappa &\rightarrow\eta_SG\kappa,\\
        N_B&\rightarrow \eta_SG\pq{N_B+\left(1-1/G\right)\left(N_V+1\right)+\frac{1-\eta_S}{\eta_SG}N_{E_1}}.        
    \end{split}
\end{equation}%
If we combine this reparameterization with an imperfect idler detection
\begin{equation}
        a'_I=\sqrt{\eta_I}a_I+\sqrt{1-\eta_I}a_{E_2},
\end{equation}%
with $\left\langle a_{E_2}^\dag{}a_{E_2}\right\rangle=N_{E_2}$, the CM of these two non-ideal modes $a'_A$ and $a'_I$ can be expressed as
\begin{equation}\label{VpAImat}
	\textbf{V}'_{AI}=
		\left(
		\begin{array}{cc}
			\pt{2N'_A+1}\textbf{I}	 &V'_{12}\textbf{R}\textbf{Z} 	\\
			V'_{12}\textbf{Z}\textbf{R}^T &\pt{2N'_I+1}\textbf{I} 	
		\end{array}
		\right),
\end{equation}%
where we call
\begin{equation}\label{psidef}
    \begin{split}
        N'_A&=\eta_SG\left[\kappa N_S+N_B+\pt{1-1/G}\pt{N_V+1}+\frac{1-\eta_S}{\eta_SG}N_{E_1}\right],\\
        V'_{12}&=2\sqrt{\eta_S\eta_IG\kappa N_S\left(N_S+1\right)},\\
        N'_I&=\eta_I\pt{N_S+\frac{1-\eta_I}{\eta_I}N_{E_2}}.\\
    \end{split}
\end{equation}%
It is worth noting how the dominance of $N'_A$ by $N_B$ in Eq.~(\ref{psidef}) suggests that excess noise from the electronics may not play a significant role.

\section{Correlation-to-displacement conversion in practice}
\label{nic-sec}
Since the procedure has been extensively discussed in Ref.~\cite{shi2022fulfilling}, we will not delve into it in this paper. By heterodyning mode $a'_A$, one obtains
\begin{equation}\label{VpIAHet}
    \begin{split}
        \textbf{V}_{I|A}^{\prime\left(Het\right)}=&\,\left(2E'+1\right)\textbf I,\\
        E'=&\,N'_I-\frac{\eta_S\eta_IG\kappa N_S\pt{N_S+1}}{N'_A+1}.
    \end{split}    
\end{equation}%
Correspondingly, with measurement result $\mathbf{\overline{x}}_\Pi=\pt{q_\Pi,\,p_\Pi}^T$, the mean of the non-ideal idler becomes
\begin{equation}
    \begin{split}
        \overline{\mathbf x}'_I&=\frac{\sqrt{\eta_S\eta_IG\kappa N_S\left(N_S+1\right)}}{N'_A+1}
        \left(
            \begin{array}{c}
			q_\Pi\cos\theta+p_\Pi\sin\theta	\\
			q_\Pi\sin\theta-p_\Pi\cos\theta
		\end{array}
        \right).
    \end{split}
\end{equation}%
With the imperfections in consideration, the distribution of the measurement outcomes is given by
\begin{equation}
    \begin{split}
        p\left(\overline{\mathbf x}_\Pi\right)&=\frac{\exp\left(-\frac{\left|\overline{\mathbf x}_\Pi\right|^2}{4\pt{N'_A+1}}\right)}{4\pi\pt{N'_A+1}},
    \end{split}
\end{equation}%
from which the distribution of $\mathcal M_m=\pt{q_{\Pi_m}+{\rm i} p_{\Pi_m}}$ can be obtained as
\begin{equation}
    p\left(\mathcal M_m\right)=\frac{\exp\left(-\frac{\left|\mathcal M_m\right|^2}{N'_A+1}\right)}{\pi\pt{N'_A+1}}.
\end{equation}%
Finally, by utlizing the displacement conditional on the heterodyne measurement result in the idler complex plane
\begin{equation}
    d_m=\frac{\sqrt{\eta_S\eta_IG\kappa N_S\left(N_S+1\right)}\ee^{\ii\theta}\mathcal M^\star_m}{N'_A+1},
\end{equation}%
we can express the total displacement of the collective idler mode at the output of the programmable beam splitter array, through a change of variables, as
\begin{equation}\label{csip}
    \begin{split}
        \left|d_T\right|^2=&\,\sum_{m=1}^M\left|d_m\right|^2=\xi\sum_{m=1}^Mz_i^2,\quad z_i\sim\mathcal N\left(0,1\right),\\
        \xi=&\,\frac{\eta_S\eta_IG\kappa N_S\left(N_S+1\right)}{
        2\pt{N'_A+1}},
    \end{split}
\end{equation}
where $\mathcal N\left(0,1\right)$ denotes a Gaussian distribution with zero mean and unit variance. In the following sections, we will make extensive use of the parameter $\xi$, which plays a critical role in our analysis. We note that $\left|d_T\right|^2$ satisfies the $\chi^2$ distribution Eq.~\eqref{PkMx}, with mean $2M\xi$ and variance $4M\xi^2$. Furthermore, Eq.~\eqref{VpIAHet} can be conveniently rephrased as $E'=N'_I-2\xi$.

\subsection{Performance limits of the conversion module in practice}
\begin{figure}[t!]
    \centering
    \includegraphics[width=\linewidth]{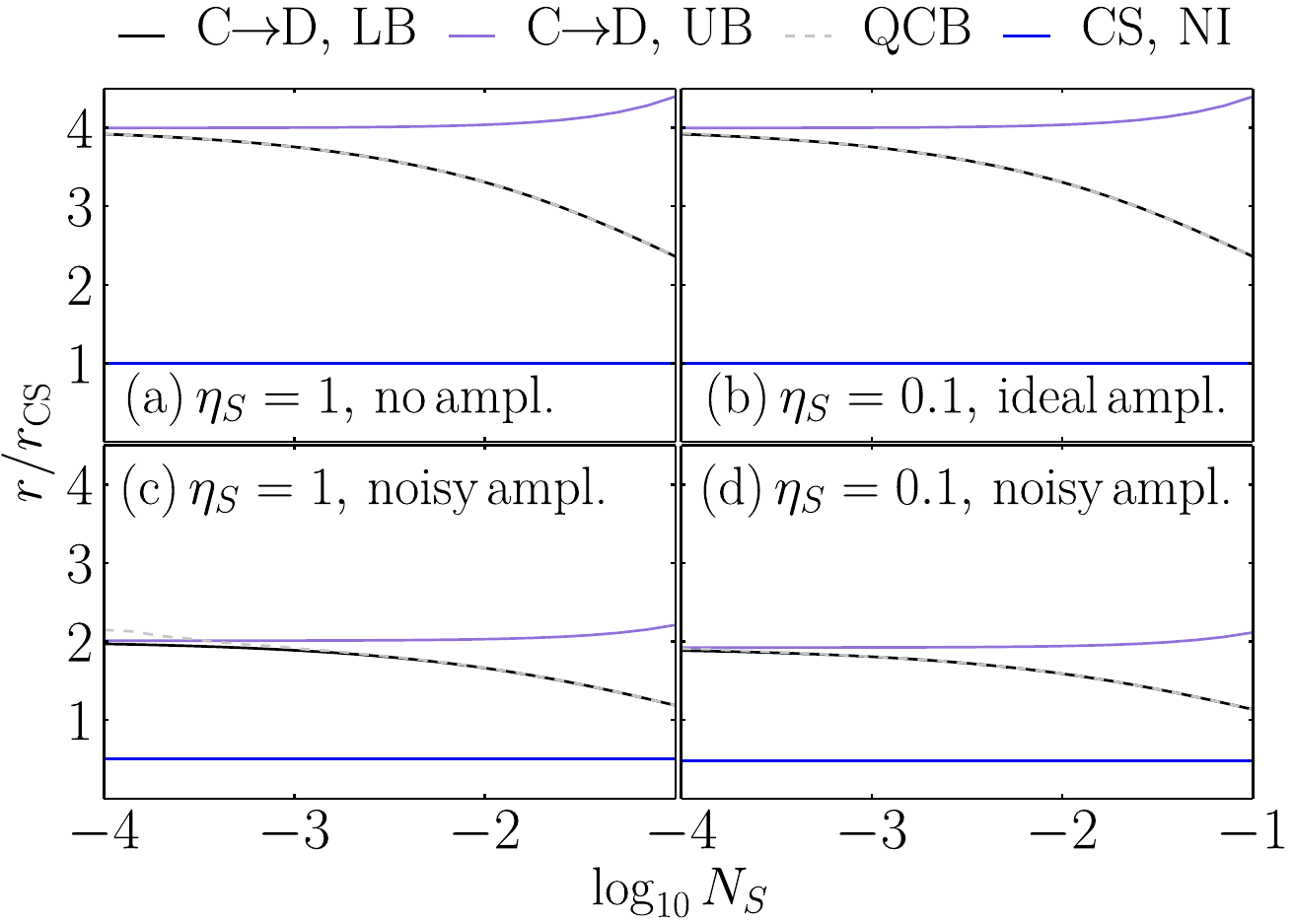}
    \caption{Black lines represent the ratio $r^{\pt{\textrm{LB}}}_{\rm C\veryshortrightarrow D}/r_\textrm{CS}$ as a function of $\log_{10}N_S$, purple ones
    $r^{\pt{\textrm{UB}}}_{\rm C\veryshortrightarrow D}/r_\textrm{CS}$, dashed gray for the QCB (see App.~\ref{QCBsec}), and blue ones $r_\textrm{CS}^{\pt{\textrm{NI}}}/r_\textrm{CS}$, where $r_{\textrm{CS}}^{\textrm{NI}}$ is obtained by applying the substitution Eq.~(\ref{chcompsub}). (a) Ideal return detection, no additional signal loss $\eta_S=1$ and therefore no amplification needed, $G=1$. (b) Lossy return detection $\eta_S=0.1$, assuming pure loss $N_{E_1}=0$. We apply quantum-limited amplification of $G=100,\,N_V=0$. (c) Ideal return detection $\eta_S=1$, and noisy amplification $G=100,\,N_V=N_B$ at room temperature. (d) Lossy return detection $\eta_S=0.1$ with noise $N_{E_1}=N_B$ at room temperature. We apply noisy amplification $G=100,\,N_V=N_B$ at room temperature. The lower bound of the $\rm C\veryshortrightarrow D$ module consistently aligns with the QCB.
    }
    \label{Fig3}
\end{figure}
The comparison between the error exponent of the $\rm C\veryshortrightarrow D$ module [see Eq.~\eqref{rCDeq} and that for the upper bound, which is within the text] and the one obtained from the Quantum Chernoff Bound (QCB) (see App.~\ref{QCBsec} for further details) can be seen in Fig.~\ref{Fig3}, showing that even in the worst case scenario of lossy amplification and imperfect detection, there is a factor of $2$ improvement compared to the classical case Eq.~\eqref{rCSeq}. Furthermore, it is worth noting that the lower bound of the conversion module consistently exhibits a close alignment with the QCB.
\begin{figure}[t!]
    \centering
    \includegraphics[width=.8\linewidth]{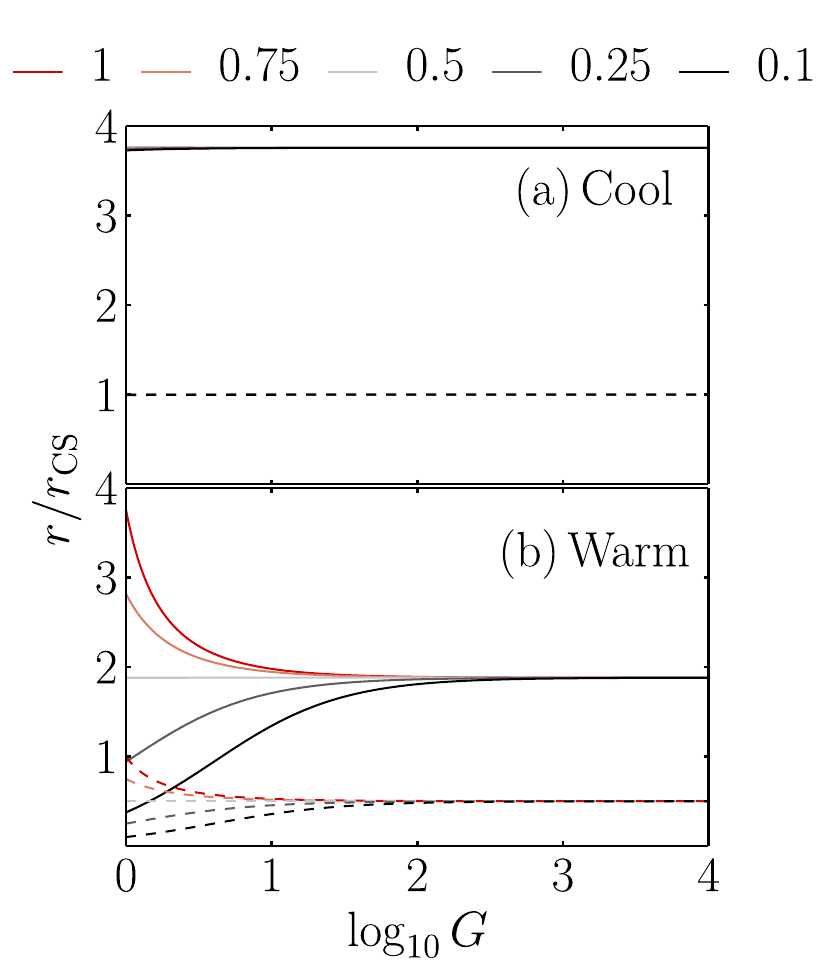}
    \caption{The impact of losses and gain on two scenarios: (a) a cool case with $N_V=N_{E_1}=0.1$ (corresponding to $T_S=100$ mK) and (b) a warm one with $N_V=N_{E_1}=N_B$ (corresponding to $T_S=300$ K). The other parameters are fixed at $N_B=1250$, $\kappa=0.01$, $N_S=10^{-3}$, and $\eta_I=1$ (i.e., we assume the idler is ideally stored). Solid lines represent the ratio $r^{\pt{\textrm{LB}}}_{\rm C\veryshortrightarrow D}/r_\textrm{CS}$ as a function of the gain $\log_{10}G$, for different values of $\eta_S$ (visible in the legend); dashed lines $r_\textrm{CS}^{\pt{\textrm{NI}}}/r_\textrm{CS}$, where $r_{\textrm{CS}}^{\textrm{NI}}$ is obtained by applying the substitution Eq.~(\ref{chcompsub}). Amplification is not necessary in a cool environment (a), but it is crucial in practical cases characterized by warm environments (b) where $\eta_S<1/2$: only through amplification can a factor of $2$ advantage be achieved.
    }
    \label{Fig4}
\end{figure}
The plots in Fig.~\ref{Fig4} provide evidence for the importance of an amplification stage in the microwave domain, where losses from detection may be challenging to overcome. It compares the ratio $r_{\rm C\veryshortrightarrow D}^{\pt{\textrm{LB}}}/r_\textrm{CS}$ with $r^{\pt{\textrm{NI}}}_\textrm{CS}/r_\textrm{CS}$ versus $\log_{10}{G}$, in two different temperature conditions (cool and warm). It can be seen that amplification is not necessary in a cool environment, but it is crucial in practical cases characterized by warm environments where $\eta_S<1/2$: only through amplification can a factor of $2$ advantage be achieved, with the emergence of an optimal value of $G$. In the later part of the paper, we will refer to the parameter setting above as either the `cool case' or the `warm case', referring to the processing temperature of the returned signal.

\subsection{Kennedy receiver}
Let us now study the performance of the $ \rm C\veryshortrightarrow D$ module in the case of a specific detection scheme of the conditional idler state. A simple idler's detection scheme is the classical Kennedy receiver, described by the set of POVMs $\Pi_0 = \left|0\right\rangle\!\left\langle0\right|$ and $\Pi_1 = \mathds{1} - \Pi_0$, where $\mathds{1}$ is the identity operator and $\left|0\right\rangle\!\left\langle0\right|$ represents the absence of a photon. The receiver operates in the on/off mode and distinguishes between the presence or absence of a photon. 

A practical approach to implement such a receiver is described in Ref.~\cite{assouly2022demonstration}, where the authors introduce a method based on a photo-current and photo-counting discriminator. While the calibration and measurement of every parameter in their system are rather complex, the basic idea is to use a dispersive qubit to read out single photons in a regime where the probability of having more than one photon is low.

We present a simple approach that provides useful insights and motivates the adoption of a Kennedy receiver, but we will not employ it for our analysis. In the limit where the number of signal photons $N_S\ll1$ is low, the receiver (neglecting experimental limitations) accurately selects $\left|0\right\rangle$ as the measurement outcome. However, the uncertainty in the decision arises from the fluctuations in the coherent state $\left|\alpha\right\rangle$. When the least probable classical situation $p_0=p_1=1/2$ is considered, the error probability can be calculated as~\cite{shi2022fulfilling}
\begin{equation}\label{pKenid}
    p_e=\frac{1}{2}\left\langle\alpha\right|\Pi_1\left|\alpha\right\rangle=\frac{1}{2}\ee^{-\left|\alpha\right|^2}
    \sim 2P_\textrm{H}
    \Rightarrow P_{K}\sim2P_{\rm C\veryshortrightarrow D},
\end{equation}%
when $\left|\alpha\right|\gg 1$ [see Eq.~(\ref{PCD})] . 

Nevertheless, the idler photon counting formula Eq.~(\ref{pKenid}) only considers the ideal case of vacuum versus coherent state. To account for deviations from this ideal scenario, we introduce a Kennedy receiver that attempts to discriminate between two differently displaced thermal states at finite $N_S$. In the $P$-representation, the two density operators to be distinguished are described by~\cite{glauber1963coherent}
\begin{equation}\label{rhoth}
    \rho_{th}\pt{\delta}=\int_{\mathbb C}\frac{d^2\beta}{\pi N_T}\,\exp\pq{-\frac{\left|\beta-\delta\right|^2}{N_T}}\left|\beta\right\rangle\!\left\langle\beta\right|,
\end{equation}%
where $\delta=\pg{0,\,\sqrt{x}}$ is the phase-space displacement, and $N_T=N'_I-\pg{0,\,2\xi}$ represents the average number of photons produced by thermal noise, with $N'_I$ and $\xi$ defined in Eqs.~(\ref{psidef}) and~(\ref{csip}), respectively. The error probability can then be calculated using the two POVMs as 
\begin{equation}\label{pe1}
    \begin{split}
        p_e&=p_0\textrm{Tr}\pq{\Pi_1\rho_{th}\pt{0}}+p_1\textrm{Tr}\pq{\Pi_0\rho_{th}\pt{\alpha}}\\
        &=p_0\pg{\mathds{1}-\textrm{Tr}\pq{\Pi_0\rho_{th}\pt{0}}}+p_1\textrm{Tr}\pq{\Pi_0\rho_{th}\pt{\alpha}},
    \end{split}
\end{equation}%
where $\textrm{Tr}\pq{\Pi_0\rho_{th}\pt{\delta}}=\exp\pt{-\frac{\left|\delta\right|^2}{N_T+1}}/\pt{N_T+1}$ [see Eq.~(E1) of Ref.~\cite{shi2020practical}]. Applied to our case, the least classical probability situation $p_0=p_1=1/2$ yields
\begin{equation}
    p_e=\frac{1}{2}\pq{1+\frac{\exp\pt{-\frac{x}{N'_I+1-2\xi}}}{N'_I+1-2\xi}-\frac{1}{N'_I+1}}.
\end{equation}%
Finally, the error probability of the Kennedy receiver is given by
\begin{equation}
    P_K=\int_0^{+\infty} dx\,P^{\pt{M}}\pt{x;\,\xi}p_e,
\end{equation}%
with $P^{\pt{M}}\pt{x;\,\xi}$ given in Eq.~(\ref{PkMx}). In other words
\begin{equation}\label{PKenfinNs}
         P_K=\frac{1}{2\pt{N'_I+1}}\pq{\pt{1+\frac{2\xi}{N'_I+1-2\xi}}^{1-M}+N'_I}.
\end{equation}%
While we have adopted the Kennedy receiver in this work, it is worth noting that further performance improvements can be achieved by optimizing the displacement amplitude and consider the improved Kennedy receiver~\cite{takeoka2008}.

\section{Performance benchmarks} \label{Perf-sec}
In order to assess the performance of the C$\veryshortrightarrow$D module, we compare it with a classical benchmark based on coherent states and homodyne detection. The error probability of homodyne detection is given by~\cite{tan2008quantum}
\begin{equation}\label{PEhomo}
    P_\textrm{E,\,homo}=\frac{1}{2}\erfc\pq{{\sqrt{\frac{\kappa M N_S}{2\pt{2N_B+1}}}}},
\end{equation}%
where ${\rm erfc}[z]=\pt{2/\sqrt{\pi}}\int_z^{+\infty}dt\,e^{-t^2}$ is the complementary error function.

\begin{figure}[t!]
    \centering
    \includegraphics[width=.9\linewidth]{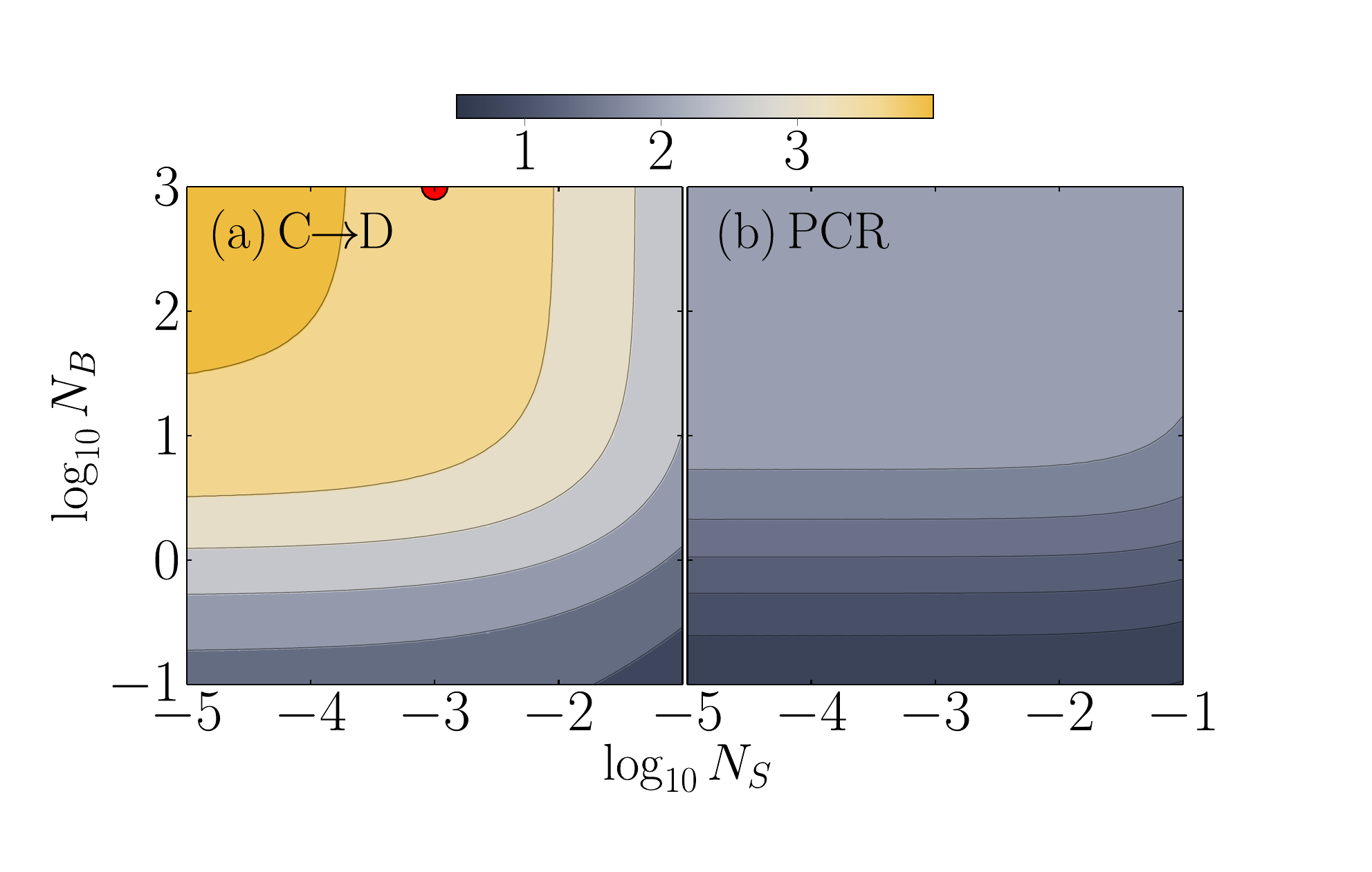}
    \caption{Comparison of the error exponent ratio $r/r_\textrm{CS}$ between the C$\veryshortrightarrow$D module and the PCR  [see Eqs.~(\ref{rCDeq}),~(\ref{PCReq}), and~(\ref{rCSeq}), respectively] as a function of $\log_{10}N_S$ and $\log_{10}N_B$. The other parameters correspond to the `cool' case and are set to: $N_V=N_{E_1}=N_{V_\textrm{PCR}}=0.1$ (corresponding to $T_S=100$ mK), $N_{E_2}=4\times 10^{-11}$ (corresponding to $T_I=10$ mK), $G=100$, $G_\textrm{PCR}=2$, $\eta_S=0.1$, and $\eta_I=0.9$. The red circle represents the parameters used in Fig.~\ref{Fig6}. The C$\veryshortrightarrow$D module possesses clear better performance, as stated by the wide yellowish areas.
    }
    \label{Fig5}
\end{figure}

Besides the classical scheme, we also benchmark with known practical receivers for QI such as the PCR scheme~\cite{shi2022fulfilling, Guha2009}, whose error probability in the QI scenario is simply given by (details can be found in App.~\ref{PCRsec})
\begin{equation}\label{PCReq}
    \begin{split}
        P_\textrm{E,PCR}&=\frac{1}{2}\erfc\pt{\sqrt{R_\textrm{PCR}M}},\\
        R_\textrm{PCR}&=\frac{\mu_1^2}{4}\left[2N'_I+\pt{G_\textrm{PCR}-1}\pt{2N'_I+1}\pt{N'_A+N'_{A,\,\kappa=0}+2}\right. \\
        &\left.+\mu_1^2/2+2G_\textrm{PCR}N_{V_\textrm{PCR}}\right]^{-1},
    \end{split}
\end{equation}%
where $\mu_1$ is given by Eqs.~(\ref{paramgausspcr}), and $G_{\rm PCR}$ and $N_{V_{\rm PCR}}$ correspond to the gain and mean number of added photons of the phase conjugator, respectively. Fig.~\ref{Fig5} shows a comparison between the performance limits of the C$\veryshortrightarrow$D module and PCR in terms of error exponents [see Eqs.~(\ref{rCDeq}) and~(\ref{PCReq}), respectively]. Although we only present the performance analysis for the cool case of return signal processing, it is noteworthy that the C$\veryshortrightarrow$D module exhibits superior performance compared to the PCR, as evidenced by a significantly larger region of parameter space with better performance, as indicated by the yellow coloration.
\begin{figure}[t!]
    \centering
    \includegraphics[width=0.8\linewidth]{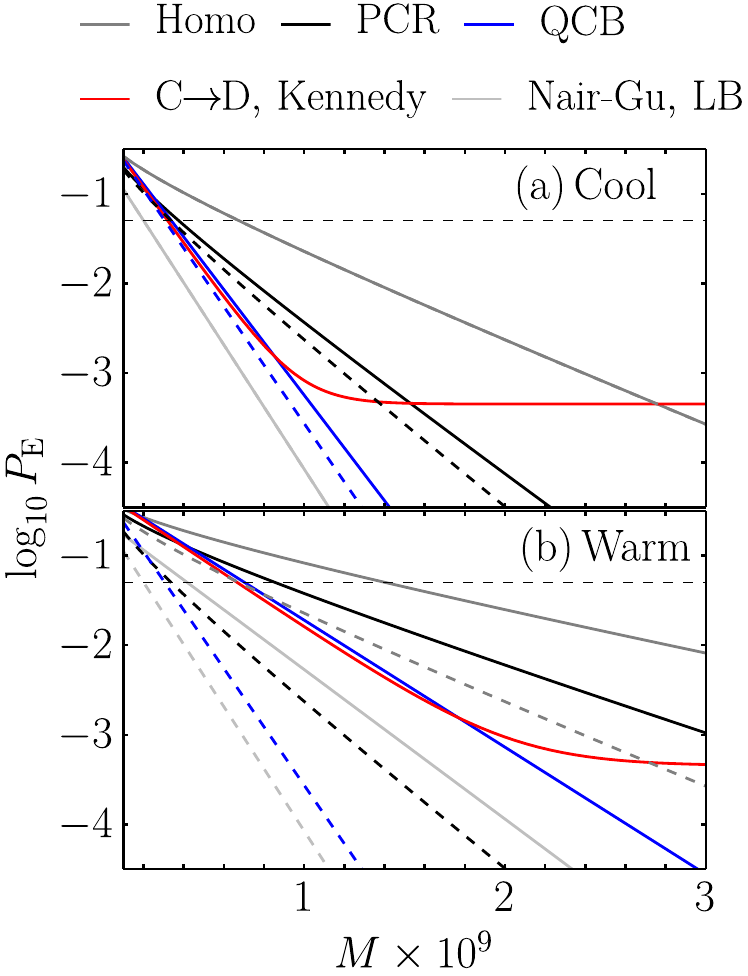}
    \caption{Error probability as a function of the number of copies $M$ in both the non-ideal (solid) and ideal (dashed) case. The non-ideal case is characterised by: $N_S=10^{-3}$, $N_B=1250$, $N_{E_2}=4\times10^{-11}$ (corresponding to $T_I=10$ mK), $\kappa=0.01$, $G=100$, $\eta_S=0.1$, $\eta_I=0.9$, and $G_\textrm{PCR}=2$. (a) Cool case with $N_V=N_{E_1}=N_{V_\textrm{PCR}}=10^{-1}$ (corresponding to $T_S=100$ mK), (b) warm one $N_V=N_{E_1}=N_{V_\textrm{PCR}}=1250$ (corresponding to $T_S=300$ K).
    Dashed lines are the performance for each solid colored curve in the ideal scenario ($\eta_S=\eta_I=1$ and no amplification $G=1$). The horizontal dashed line marks $P_\textrm{E,\,homo}=0.05$.
    }
    \label{Fig6}
\end{figure}
The scaling of major error probabilities with the number of copies $M$ is shown in Fig.~\ref{Fig6}, for both the warm and cool cases. Note that the parameter setting of Fig.~\ref{Fig6} corresponds to the red dot in Fig.~\ref{Fig5}. Specifically, we focus on the performance of the C$\veryshortrightarrow$D module with Kennedy receiver (red lines), which is almost comparable to that of the QCB (blue) and outperforms any other practical scheme considered. The saturation of the C$\veryshortrightarrow$D performance is due to the on-off detection of Kennedy receiver, as we will resolve in Sec.~\ref{sec:number_resolving}. We also present the comparison to the Nair-Gu lower bound~\cite{Nair:20} (light gray), which shows similar scaling of the QCB. In Fig.~\ref{Fig5}, the dashed curves are the performance curves of the receivers assuming all equipment become ideal, instead the solid curves where imperfections are considered (the same color coding of the curves are adopted for both dashed and solid, as indicated by the legend).
To provide a comparison between the C$\veryshortrightarrow$D module equipped with an on/off Kennedy receiver and the PCR, Fig.~\ref{Fig7} presents the error probability ratio $\log_{10}\pt{P_E/P_\textrm{E,\,homo}}$ for the cool case, where $M$ is chosen such that the homodyne error probability is fixed at $P_\textrm{E,\,homo}=0.05$. As shown by the wide dark area, the C$\veryshortrightarrow$D module clearly outperforms the PCR in the $N_B \gg 1$, $N_S \ll 1$ parameter regime.

\section{Enhanced performance with number-resolving detection}
\label{sec:number_resolving}
So far we have adopted the Kennedy receiver with on-off detection, which leads to the saturation of error probability (red lines) in Fig.~\ref{Fig6} at large $M$. To obtain better performance, in this section we generalize the Kennedy receiver to a photon number resolving detector (PNRD) on the idler.

As already analyzed, the decision between the presence or absence of the target is equivalent to discriminating between two states of the final idler mode after the beamsplitter array: the thermal state $\rho_{0,\,N_I'}$ when the target is absent, and the displaced thermal state $\rho_{\sqrt{x},\,E'}$ when it is present. Recall that $N_I'$ is defined by Eq.~(\ref{psidef}), $E'$ by Eq.~(\ref{VpIAHet}), and $x$ is a random variable associated with the results of $M$ heterodyne measurements on the return modes, distributed according to Eq.~(\ref{PkMx}), with $\xi$ given by Eq.~(\ref{csip}). With a PNRD detection, we can now compare the photon number probability distributions for the two hypotheses: $p_n^{\pt{0}}=\left\langle n \right |\rho_{0,\,N_I'}\left |n\right\rangle$ and $p_n^{\pt{1}}(x)=\left\langle n \right |\rho_{\sqrt{x},\,E'}\left |n\right\rangle$. The presence of the target is declared when the outcome of the photon number measurement is greater than a predetermined threshold value, $n\geq n_D\geq 1$. 

To prepare our analyses for the ROC curve, we consider 
the false alarm probability $P_F$ and the detection probability $P_D$ for a fixed decision threshold $n_D$ as
\begin{eqnarray}
    P_F&=&\sum_{n=n_D}^{+\infty} \left\langle n \right |\rho_{0,\,N_I'}\left |n\right\rangle, \label{pdpf1}\\
    P_D&=&\sum_{n=n_D}^{+\infty} \int_0^{+\infty}dx\,P^{\pt{M}}\pt{x;\,\xi}\left\langle n \right |\rho_{\sqrt{x},\,E'}\left |n\right\rangle, \label{pdpf2}
\end{eqnarray}%
where we average over the random variable $x$. 

\begin{figure}[t!]
    \centering
    \includegraphics[width=\linewidth]{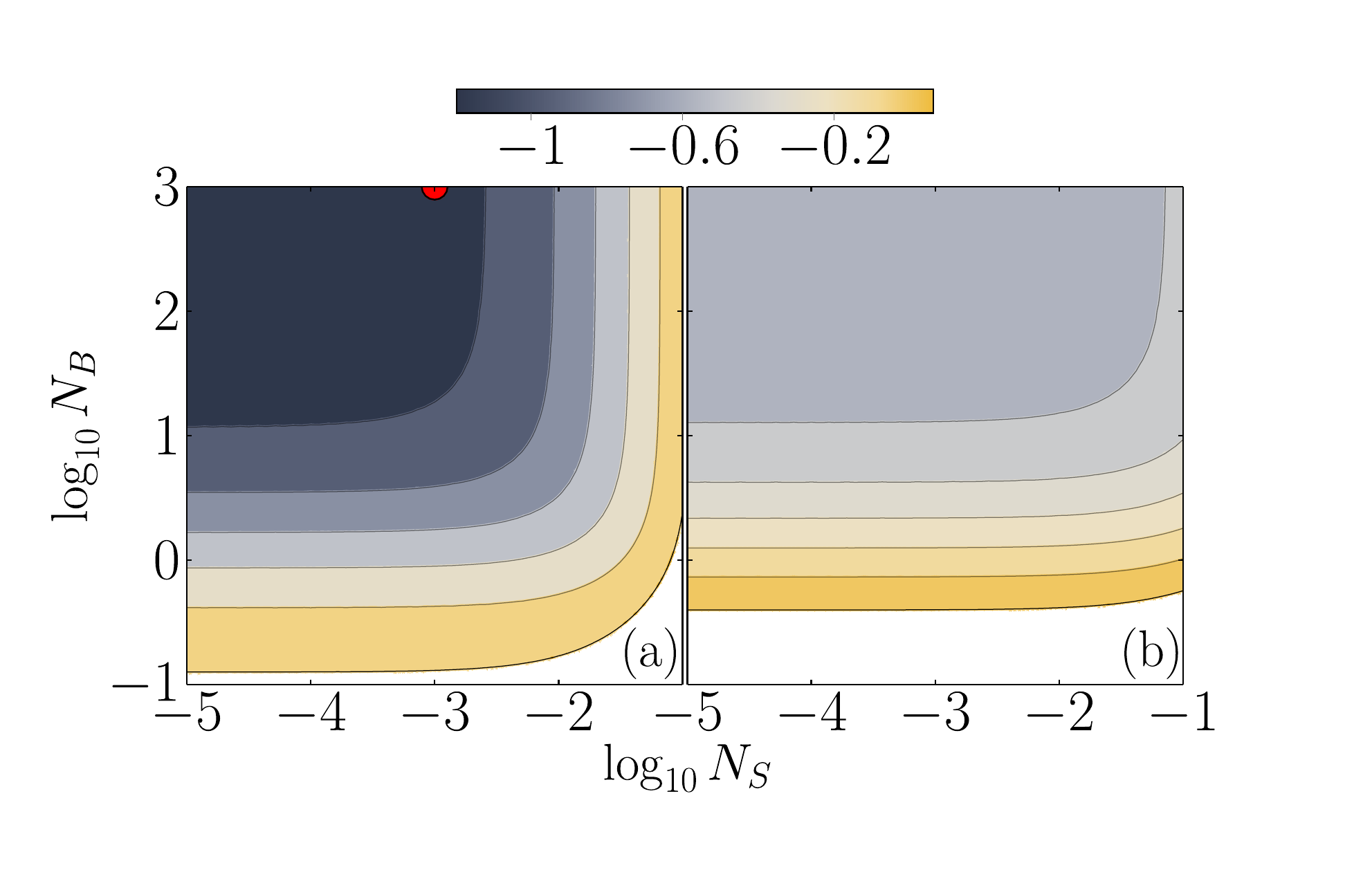}
    \caption{Comparison based on the error probability ratio $\log_{10}\pt{P_E/P_\textrm{E,\,homo}}$ between (a) the C$\veryshortrightarrow$D module (equipped with an on/off Kennedy receiver) and (b) the PCR  [see Eqs.~(\ref{PKenfinNs}),~(\ref{PCReq}), and~(\ref{PEhomo}), respectively] vs. $\log_{10}N_S$ and $\log_{10}N_B$. The value of $M$ is selected to set $P_\textrm{E,\,homo}=0.05$. The other parameters correspond to the `cool' case and are: $N_V=N_{E_1}=N_{V_\textrm{PCR}}=0.1$, $N_{E_2}=4\times 10^{-11}$, $G=100$, $G_\textrm{PCR}=2$, $\eta_S=0.1$, and $\eta_I=0.9$. The red circle indicates the parameters used in Fig.~\ref{Fig6}. As shown by the wide dark area, the C$\veryshortrightarrow$D module outperforms the PCR.
    }
    \label{Fig7}
\end{figure}

The evaluation of $P_F$ is simple and one has
\begin{equation}\label{pf-fin}
    P_F\pt{n_D}=\left(\frac{N_I'}{N_I'+1}\right)^{n_D},
\end{equation}
while that of $P_D$ is more involved. We start by using the following result for the photon statistics of a displaced thermal state for a given $x$~\cite{Lachs1965,Marian1993}
\begin{equation} \label{pdx}
        p_n^{\pt{1}}(x) =\frac{\exp\left(-\frac{x}{E'+1}\right)}{E'+1}\left(\frac{E'}{E'+1}\right)^n L_n\left[-\frac{x}{E'(E'+1)}\right],
\end{equation}
where $L_n\pq{\cdot}$ is the $n$-th Laguerre polynomial. Next, one can perform the average over the probability distribution Eq.~(\ref{PkMx}) to obtain the average photon number probability distribution when the target is present
\begin{equation}\label{pdaver}
    \begin{split}
        \bar{p}_n^{\pt{1}}\pt{M;\,\xi}&=\frac{(E'+1)^{M-n-1}E'^n}{(E'+1+2\xi)^M}\\
        &\times{}_2F_1\pq{M,\,-n,\,1,\,-\frac{2 \xi}{E'(E'+1+2\xi)}},
    \end{split}
\end{equation}
where $_2F_1\pt{a,\,b,\,c,\,z}$ is the Gaussian hypergeometric function. Consequently, the detection probability $P_D\pt{n_D}$ can be exactly determined as
\begin{equation}\label{pdfinale}
    P_D\pt{n_D}=1-\sum_{n=0}^{n_D-1} \bar{p}_n^{\pt{1}}\pt{M;\,\xi}. 
\end{equation}

\subsection{Bayesian error probability}
 To begin with, we consider the symmetric error
$P_E=(P_F+1-P_D)/2$ and evaluate the performance. Here the results are similar to that of Ref.~\cite{https://doi.org/10.48550/arxiv.2212.08190}. This is because, given the choice of photon counting, random phase does not change the performance anymore. From Eqs.~\eqref{pdaver} and~\eqref{pf-fin}, we have the error probability $P_E$ as a function of the threshold $n_D$.
We compare this optimal decision strategy with a variable threshold $n_D$, and quantify the error of probability using
\begin{equation}\label{PCDnD}
    P_{\rm C\veryshortrightarrow D}^{\pt{n_D}}=\frac{1}{2}\pq{1-\sum_{n=0}^{n_D-1}\gamma_{n}\pt{2M;\,\xi}},
\end{equation}
where the function
\begin{equation}
    \begin{split}
        &\gamma_{n}\pt{M;\,\xi}=\frac{N^{\prime n}_I}{\pt{N'_I+1}^{n+1}}\\
        &-\frac{(E'+1)^{M-n-1}E'^n}{(E'+1+2\xi)^M} {}_2F_1\pq{M,\,-n,\,1,\,-\frac{2 \xi}{E'(E'+1+2\xi)}}.
    \end{split}
\end{equation}
Although the above equation is exact, to enable efficient numerical evaluation in all parameter region of interest, we further make an approximation at the $M\gg1$ limit and obtain 
 \begin{equation}
    \begin{split}
        \gamma_{n}\pt{M;\,\xi}&\simeq \frac{N^{\prime n}_I}{\pt{N'_I+1}^{n+1}}\\
        &-\frac{E^{\prime n}}{\pt{E^\prime+1}^{n+1}}\ee^{-2M\xi/E'}{}_1{F}_1\pq{n+1,\,1,\,\frac{2M\xi}{E'\pt{E'+1}}}.
    \end{split}
\end{equation}
The precision of such an approximation is sufficient for our evaluation, as verified in Ref.~\cite{chen2022}.
 The optimal performance is then given by a minimization of the error probability over the threshold $n_D$,
\begin{equation}\label{PCDopt}
    P_{\rm C\veryshortrightarrow D}^{\pt{opt}}=\min_{n_D\ge 1} P_{\rm C\veryshortrightarrow D}^{\pt{n_D}}.
\end{equation}
Note that $P_{\rm C\veryshortrightarrow D}^{\pt{1}}\equiv P_K$ [see Eq.~\eqref{PKenfinNs}], as expected.

\begin{figure}[t!]
    \centering
    \includegraphics[width=0.9\linewidth]{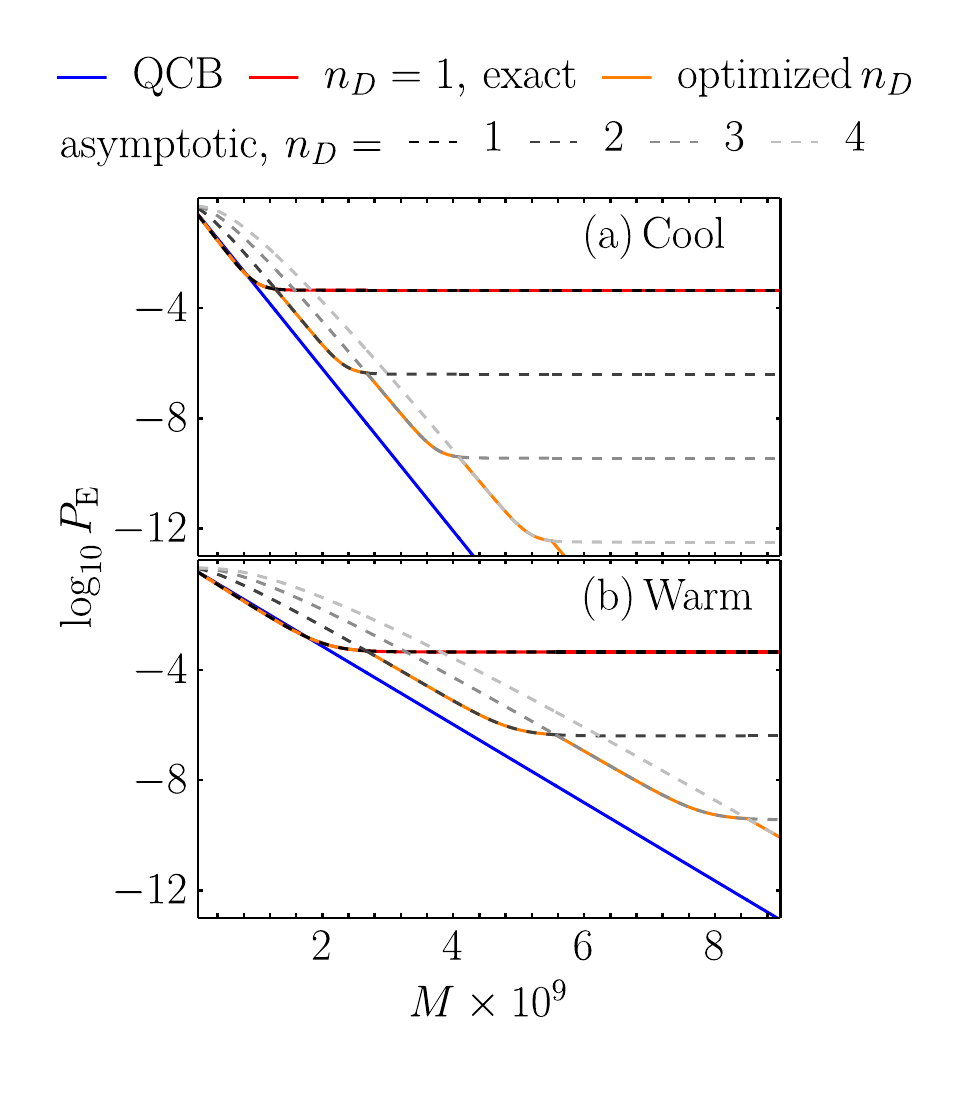}
    \caption{The saturation of the red line in Fig.~\ref{Fig6} suggests an improvement, following the lines of Ref.~\cite{https://doi.org/10.48550/arxiv.2212.08190}, where a variable threshold decision strategy approach has been used for asymptotic analysis. The red curve reproduces the usual Kennedy receiver corresponding to the fixed threshold $n_D =1$. The dashed grey lines corresponds to the case of fixed, increasing values of $n_D$. The orange line gives the optimized result in which $n_D$ is adjusted according to $M$, and therefore to the two states to be discriminated. This latter approach yields results comparable to those of the non-ideal QCB (blue). Parameter values are the same as those of Fig.~\ref{Fig6}.
    }
    \label{Fig8}
\end{figure}

Fig.~\ref{Fig8} shows the results of our analysis, using the same parameter values as in Fig.~\ref{Fig6}. We observe that the optimized approach (orange) produces results that are comparable to those of the non-ideal QCB (blue, see also App.~\ref{QCBsec}). Specifically, the irregular trend in the data is well described by a variable threshold decision strategy approach, which is represented by the dashed lines in the figure. Our findings suggest that the optimized approach can effectively discriminate between the two states of interest, even in the presence of noise and other imperfections.

\subsection{Receiver operating characteristic}\label{roc-sec}

\subsubsection{Conversion module and photon-number resolving detector}
Let us now analyse the performance of the C$\veryshortrightarrow$D module within the Neyman-Pearson framework, using ROC curves. In this approach, a chosen false alarm probability $P_F$ is fixed, and the goal is to maximize the detection probability $P_D$. 
By gradually reducing the threshold value $n_D$ from a high (ideally infinite) value to zero, a concave ROC curve can be obtained, plotting $P_D$ versus $P_F$, starting from $P_F = P_D = 0$ and ending at $P_F = P_D = 1$.

To gain a clearer understanding of the behavior of the ROC curve, we derive an analytical expression based on a Gaussian approximation. When $x \gg 1$, the probability distribution $p_n^{\pt{1}}(x)$ Eq.~(\ref{pdx}) can be represented by a Gaussian distribution with mean $\left\langle n(x) \right\rangle =E'+x$, and variance $\sigma_n^2(x)=\left\langle n^2(x) \right\rangle -\left\langle n(x) \right\rangle^2 =E'^2+E'+x\pt{2E'+1}$. As a result, in this limit, the average probability distribution Eq.~(\ref{pdaver}) can also be approximated by a Gaussian distribution with properly averaged mean and variance, and we have
\begin{equation}\label{pdaverappro}
        \bar{p}_n^{\pt{1}}\pt{M;\,\xi}\sim \frac{1}{\sqrt{2\pi \sigma_n^2}} \exp\pq{-\frac{\pt{n-\bar{n}}^2}{2 \sigma_n^2}},
\end{equation}
with 
\begin{eqnarray}
    \bar{n}&=&E'+\bar{x} \nonumber\\
    &=&E'+2 M \xi, \label{meanapp}\\
    \sigma_n^2&=&E'^2+E'+\bar{x}\pt{2E'+1}+\sigma_x^2 \nonumber\\
    &=& E'^2+E'+2M\xi\pt{2\xi+2E'+1},\label{varapp}
\end{eqnarray}%
taking into account that the distribution Eq.~(\ref{PkMx}) has mean $\bar{x}=2M\xi$, and variance $\sigma_x^2=4M\xi^2$. A necessary condition for the validity of such a Gaussian treatment is that $\bar{x}=2M\xi \gg 1$. By using the Gaussian approximation Eq.~(\ref{pdaverappro}), and eliminating the threshold $n_D$ with the aid of Eq.~(\ref{pf-fin}), one gets the following approximate expression for the ROC curve of the C$\veryshortrightarrow$D module
\begin{equation}\label{roc-appro}
    P_D\pt{P_F} \sim \frac{1}{2} \erfc\pq{\frac{1}{\sigma_n \sqrt{2}}\left(\frac{\log P_F}{\log\left(\frac{N_I'}{N_I'+1}\right)}-\bar{n}\right)}.
\end{equation}
This approximation provides a satisfactory description of the ROC curves for moderate values of $P_D$ and $P_F$ as long as $ 2 M\xi > 1$. Although the average probability distribution $\bar{p}_n^{\pt{1}}\pt{M;\,\xi}$ resembles a Gaussian distribution around the peak centered at its average value, it decays exponentially, not Gaussianly, for $P_F \to 0 \Rightarrow P_D \to 0$, i.e., $n_D \to \infty$. As a result, Eq.~(\ref{roc-appro}) tends to underestimate the value of $P_D$ for high threshold values $n_D$.

\subsubsection{The ROC curve in the case of the PCR}\label{roc_pcr_sec}
As discussed in App.~\ref{PCRsec} (see also Ref.~\cite{sorelli}), when $M\gg 1$, the photo-count difference of the PCR, $N=N_+-N_-$, according to the central limit theorem, follows a Gaussian distribution with a probability density for the two hypotheses
\begin{equation}
    P_{N\vert H_{0/1}}\pt{n\vert H_{0/1}}=\frac{\exp\pq{{-\frac{\pt{n-M \mu_{0/1}} ^{2}}{2M\sigma_{0/1}^{2}}}}}{\sqrt{2\pi M\sigma_{0/1}^{2}}},
\end{equation}
where the two mean values $\mu_{0/1}$ and the two variances $\sigma_{0/1}^{2}$ are given by Eqs.~(\ref{paramgausspcr}).

The discrimination between two Gaussian distributions with different means and variances can be obtained by using the extended van Trees approximation~\cite{shapiro1999}, and it can be expressed in terms of the auxiliary function
\begin{equation}\label{mus}
    \begin{split}
        \mu(s)=\ln&\left\{\frac{\sigma_{1}^{1-s}\sigma_{0}^{s}}{\sqrt{s\sigma_{0}^{2}+\pt{1-s}\sigma_{1}^{2}}}\right.\\
        &\left.\times
     \exp\pg{-\frac{M\pt{\mu_{0}-\mu_{1}}^{2}s\pt{1-s}}{2\pq{s\sigma_{0}^{2}+\pt{1-s}\sigma_{1}^{2}}}} \right\}, 
    \end{split}
\end{equation}
where $s$ is a threshold parameter. The false alarm and detection probabilities are then respectively given by
\begin{equation}\label{pfpcr}
    \begin{split}
        P_{F}&=\frac{1}{2}{\rm erfc}\left[ s\sqrt{\frac{\ddot{\mu}(s)}{2}}\right],  \\
        P_{D}&= 1-\frac{1}{2}{\rm erfc}\left[\pt{1- s}\sqrt{\frac{\ddot{\mu}(s)}{2}}\right],
    \end{split}
\end{equation}
where $\ddot{\mu}(s)\equiv d^2\mu/ds^2$.

However, one can get a simpler and clearer expression by taking into consideration that the variances for the two hypothesis, $\sigma^2_{0}$ and $\sigma^2_{1}$, are nearly identical for the typical parameter values in a microwave QI experiment, that is, when $\kappa \ll 1$, $N_S \ll 1$, and $N_B \gg 1$. In fact, Eqs.~(\ref{paramgausspcr}) give
\begin{equation}
    \begin{split}
       &\frac{\sigma_1^2-\sigma_0^2}{\sigma_0^2}=\eta_S G\pt{G_\textrm{PCR}-1}\kappa N_S\pq{2N'_I+1+2 \eta_I\pt{N_S+1}}\\
       &\times\pq{N'_I+\pt{G_\textrm{PCR}-1}\pt{2N'_I+1}\pt{N'_{A,\,\kappa=0}+1}+G_\textrm{PCR}N_{V_\textrm{PCR}}}^{-1},
    \end{split}
\end{equation}
which scales as $\kappa N_S/N_B \ll 1$ when $N_B \gg N_S$. As a result, one has $\ddot{\mu}(s)= M \mu_1^2/\sigma_1^2 \equiv d_\textrm{PCR}^2$ in Eqs.~(\ref{pfpcr}), which can be rewritten as 
\begin{equation}\label{pfpcr2}
    \begin{split}
        P_{F}&=\frac{1}{2} {\rm erfc}\left[\frac{1}{\sqrt{2}}\left( \frac{\ln\eta}{d_\textrm{PCR}}+\frac{d_\textrm{PCR}}{2}\right)  \right], \\
        P_{D}&=\frac{1}{2}{\rm erfc}\left[\frac{1}{\sqrt{2}}\left( \frac{\ln\eta}{d_{PCR}}-\frac{d_\textrm{PCR}}{2}\right)\right],
    \end{split}
\end{equation}
where we introduce the new threshold parametrization as $\ln \eta = \pt{s-1/2} d_\textrm{PCR}^2$. By eliminating this threshold parameter, the analytical expression of the ROC curve for the PCR can be obtained as
\begin{equation}\label{ROCpcr}
    P_{D}=\frac{1}{2}{\rm erfc}\left[{\rm erfc}^{-1}\pt{2P_F}-\frac{d_\textrm{PCR}}{\sqrt{2}}\right],
\end{equation}
where ${\rm erfc}^{-1}(z)$ is the inverse of the complementary error function. We notice that the ROC curve for the PCR is analytically identical to that of the optimal classical benchmark of using coherent states and homodyne detection. Both have the same form as in Eq.~(\ref{ROCpcr}), but the replacement $d_\textrm{PCR}\to d_\textrm{CS}=2\sqrt{M\kappa N_S/\pt{2N_B+1}}$~\cite{sorelli}.

\begin{figure}[t!]
\centering
\includegraphics[width=.8\linewidth]{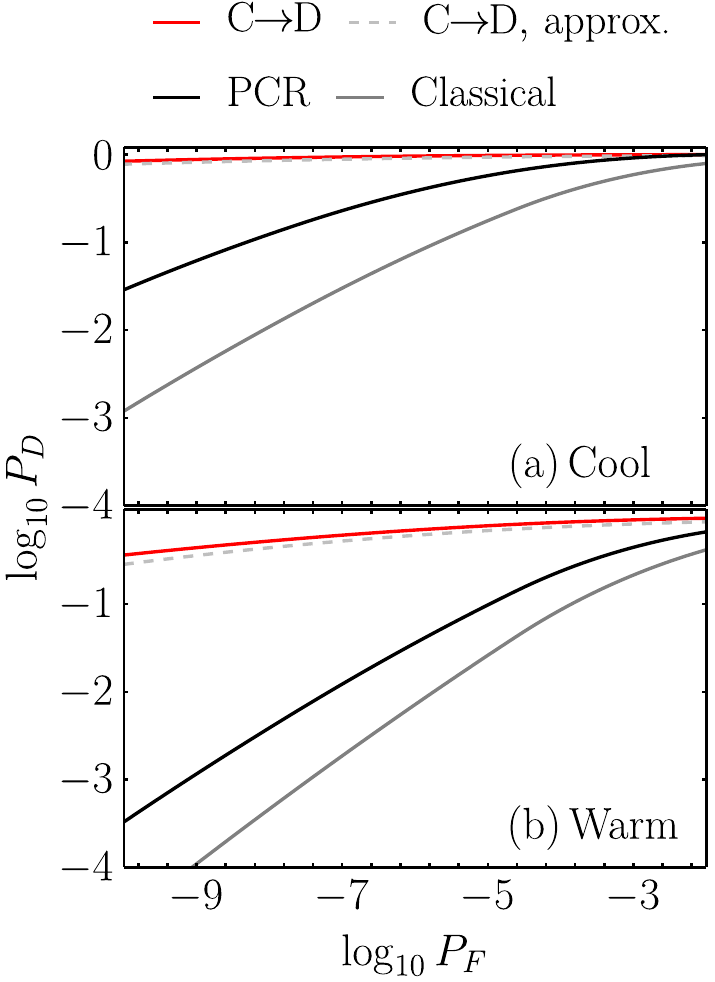}
\caption{Comparison of ROC curves. The red line shows the performance of the C$\veryshortrightarrow$D module with a PNRD; the black one that of the PCR given by Eq.~\eqref{ROCpcr}; the dashed light gray line depicts the performance of the Gaussian approximation of Eq.~\eqref{roc-appro}; the full dark grey line gives the non-ideal classical benchmark [using Eq.~\eqref{ROCpcr} with $d_\textrm{PCR}\to d_\textrm{CS}$ plus Eq.~\eqref{chcompsub}]. The parameters used are the same as in Fig.~\ref{Fig5} and~\ref{Fig7} (indicated by the red dots there), with $M=69\times10^7$.
}
\label{Fig9}
\end{figure}

Fig.~\ref{Fig9} presents the behavior of the ROC curve for the C$\veryshortrightarrow$D module in both warm and cool cases, considering losses and amplification in the detection scheme. The results are compared to the corresponding Gaussian approximation Eq.~\eqref{roc-appro}, the PCR Eq.~\eqref{ROCpcr}, and the non-ideal classical benchmark [using Eq.~\eqref{ROCpcr} with $d_\textrm{PCR}\to d_\textrm{CS}$ and Eq.~\eqref{chcompsub}], all obtained under the same experimental conditions. 

When the Neyman-Pearson decision strategy is considered, it can be observed that the C$\veryshortrightarrow$D module exhibits excellent performance in both the cool and warm cases. In particular, its ROC curve is significantly larger than those obtained with the PCR and the classical approach for the same experimental conditions.

\section{Conclusions}\label{concl}
In conclusion, this work analyzed how experimental imperfections can be mitigated using correlation-to-displacement conversion-based receivers, and our results showed that amplification on the return signals can effectively compensate for additional loss in the heterodyne detection. We also employed a Kennedy receiver for idler detection conditioned on heterodyne and demonstrated that in the ideal case, such a scheme has the optimal error exponent. In practical scenarios, such a receiver still provides quantum advantages over classical optimal schemes and outperforms other known practical receivers for quantum illumination. Overall, our findings illustrate the feasibility of practical microwave quantum illumination systems that can overcome experimental imperfections and offer quantum advantages for target detection in noisy environments. These insights can inform the development of future quantum illumination systems and contribute to the advancement of quantum sensing technologies.

\begin{acknowledgments}
J. A., T. L., and D. V. acknowledge the support of PNRR MUR project PE0000023-NQSTI (Italy), and of the European Union Horizon 2020 Programme for Research and Innovation through the Project No. 862644 FET-Open QUARTET. H. S. and Q. Z. acknowledge the support of National Science Foundation (NSF) CAREER Award CCF-2142882, NSF Engineering Research Center for Quantum Networks Grant No. 1941583, and Cisco Systems, Inc.. J. A. also acknowledges support from University of Southern California (USC) and Cisco Systems, Inc. during his visit to USC.
\end{acknowledgments}

\appendix

\section{Quantum Chernoff bound}\label{QCBsec}
The Quantum Chernoff Bound (QCB) is a powerful tool for determining an upper bound to the Helstrom limit $P_\textrm{H}$~\cite{PhysRevLett.84.2722, PhysRevLett.84.2726, Pirandola_2008, weedbrook2012gaussian}. It is particularly useful for an ensemble of Gaussian states $\{\rho_h\}$, where it can be efficiently computed using symplectic decomposition. In our specific case of discriminating between $\Phi_{0,\,0}$ and $\Phi_{\kappa,\,0}$, this corresponds to the discrimination of $\pg{\rho_h}_{h=0}^\kappa$, with mean $\overline{\mathbf x}_h=0$ and CM $\textbf{V}_{AI}^{\prime\pt{h}}$ [see Eq.~(\ref{VpAImat}) with $\theta=0$]. Indeed, the matrix $\textbf{V}_{AI}^{\prime\pt{h}}$ can be denoted as
\begin{equation}
    \textbf{V}_{AI}^{\prime\pt{h}}=
        \left(
        \begin{array}{cc}
            a_h\textbf{I}	 &c_h\textbf{Z}\\
            c_h\textbf{Z} &b\textbf{I}, 	
        \end{array}
        \right),
\end{equation}%
with $a_h=2N'_A+1$, $b=2N'_I+1$, and $c_h=V^\prime_{12}$. Its symplectic eigenspectrum is then given by
\begin{equation}
    \nu_\pm^{\pt{h}}=\frac{\sqrt{y_h}\pm\pt{b-a_h}}{2},
\end{equation}
where $y_h=\pt{a_h+b}^2-4c_h^2$, and with the symplectic matrix $\textbf{S}_h$ described by
\begin{equation}
    \textbf{S}_h=
        \left(
        \begin{array}{cc}
            \omega_+^{\pt{h}}\textbf{I}	 &\omega_-^{\pt{h}}\textbf{Z}\\
            \omega_-^{\pt{h}}\textbf{Z} &\omega_+^{\pt{h}}\textbf{I}, 	
        \end{array}
        \right),\quad\omega_\pm^{\pt{h}}=\sqrt{\frac{a_h+b\pm\sqrt{y_h}}{2\sqrt{y_h}}}.
\end{equation}%
In this regard, the QCB is simply expressed as
\begin{equation}\label{PQCBeq}
    P_\textrm{QCB}=\frac{1}{2}\pt{\inf_{s\in[0,\,1]}\overline{Q}_s}^M,
\end{equation}%
where $\overline{Q}_s=4\,\textrm{det}\pt{\boldsymbol{\Sigma}_s}^{-1/2}\prod_{j=1}^2G_s\pt{\nu_j^{(0)}}G_{1-s}\pt{\nu_j^{(\kappa)}}$, $\boldsymbol{\Sigma}_s=\widetilde{\textbf{V}}_0(s)+\widetilde{\textbf{V}}_\kappa(1-s)$, and having defined
\begin{equation}\label{QCB_theo}
    \begin{split}
        \widetilde{\textbf{V}}_h(s)&=\textbf{S}_h\pq{\bigoplus_{j=1}^2\Lambda_s\pt{\nu_j^{(h)}}\textbf I}\textbf{S}_h^T,\\
        G_s(x)&=\frac{2^s}{\pt{x+1}^s-\pt{x-1}^s},\\
        \Lambda_s(x)&=\frac{\pt{x+1}^s+\pt{x-1}^s}{\pt{x+1}^s-\pt{x-1}^s}.
    \end{split}
\end{equation}%

\section{Phase-conjugate receiver}\label{PCRsec}
To ensure clarity and avoid confusion for the reader, we reintroduce the hat notation in this section, to distinguish between an operator $\hat O$ and its corresponding mean value $O=\left\langle\hat O\right\rangle$.

In a PCR the $\hat a'_A$ modes are phase-conjugated according to the following transformation
\begin{equation}
    \hat a_C=\sqrt{G_\textrm{PCR}}\hat a_{V_\textrm{PCR}}+\sqrt{G_\textrm{PCR}-1}\hat a_A^{\prime\dag{}},
\end{equation}%
where $G_{\rm PCR}$ is the gain and $\hat a_{V_\textrm{PCR}}$ is the annihilation operator of the noise entering the unused port of the PCR. The conjugated modes are then recombined on a balanced beamsplitter with the non-ideal idler mode $\hat a'_I$ as $\hat a_\pm=\pt{\hat a_C\pm \hat a'_I}/\sqrt{2}$, that is
\begin{equation}
    \hat a_\pm=\frac{1}{\sqrt{2}}\pt{\sqrt{G_\textrm{PCR}}\hat a_{V_\textrm{PCR}}+\sqrt{G_\textrm{PCR}-1}\hat a_A^{\prime\dag{}}\pm \hat a'_I}. 
\end{equation}
In the following analysis, we will not consider terms whose mean value $\left\langle\,\cdot\,\right\rangle$ is null, such as those linear in $\hat a_{V_\textrm{PCR}}$. Similarly, we will group together terms whose mean value $\left\langle\,\cdot\,\right\rangle$ is equal, that is, $\left\langle \hat a'_A\hat a'_I\right\rangle=\left\langle \hat a_I^{\prime\dag{}}\hat a_A^{\prime\dag{}}\right\rangle$. That said, the photon numbers at the output of the beamsplitter can be expressed as
\begin{equation}
    \begin{split}
        \hat N_\pm&=\hat a_\pm^\dag{}\hat a_\pm\\
        &=\frac{1}{2}\pq{\hat N_C\pm2\sqrt{G_\textrm{PCR}-1}\hat a'_A\hat a'_I+\hat N'_I},
    \end{split} 
\end{equation}
where
\begin{equation}\label{NCNVPCR}
    \begin{split}
        \hat N_C&=\hat a_C^\dag{}\hat a_C\\
        &=G_\textrm{PCR}\hat N_{V_\textrm{PCR}}+\pt{G_\textrm{PCR}-1}\pt{\hat N'_A+1},\\
        \hat N_{V_\textrm{PCR}}&=\hat a_{V_\textrm{PCR}}^\dag{}\hat a_{V_\textrm{PCR}}.
    \end{split}
\end{equation}
When $M\gg 1$, the photo-count difference $N=N_+-N_-$, according to the central limit theorem, follows a Gaussian distribution with mean $M\mu$, where 
\begin{equation}\label{mu_pcr_eq}
    \mu=\sqrt{G_\textrm{PCR}-1}V'_{12},
\end{equation}
and variance $M\sigma^2$, with
\begin{equation}\label{sigma_pcr_eq}
    \begin{split}
        \sigma^2&=N_+\pt{N_++1}+N_-\pt{N_-+1}-\pt{N_C-N'_I}^2/2\\
        &=G_\textrm{PCR}N_{V_\textrm{PCR}}+\pt{G_\textrm{PCR}-1}\pt{N'_A+1}+N'_I\\
        &+2\pt{G_\textrm{PCR}-1}\pt{N'_A+1}N'_I+\pt{G_\textrm{PCR}-1}V_{12}^{\prime2}/2.
    \end{split}
\end{equation}%
The values of the mean $\mu$ and the variance $\sigma^2$ are influenced by both the off-diagonal CM element $V_{12}'$ and $N'_A$, which vary depending on whether the target is present ($H_1$) or absent ($H_0$). Using the Gaussian approximation, in the QI scenario, the error probability is simply given by
\begin{equation}
        P_\textrm{E,PCR}=\frac{1}{2}\erfc\pt{\sqrt{R_\textrm{PCR}M}},\quad
        R_\textrm{PCR}=\frac{\pt{\mu_1-\mu_0}^2}{4\pt{\sigma_1^2+\sigma_0^2}},
\end{equation}%
where the mean and variance for the two hypotheses are given by
\begin{equation}
    \begin{split}
        \mu_0&=0,\\
        \mu_1&=2\sqrt{\eta_S\eta_IG\pt{G_\textrm{PCR}-1}\kappa N_S\pt{N_S+1}},\\ \sigma_0^2&=N'_I+\pt{G_\textrm{PCR}-1}\pt{2N'_I+1}\pt{N'_{A,\,\kappa=0}+1}+G_\textrm{PCR}N_{V_\textrm{PCR}},\\        \sigma_1^2&=N'_I+\pt{G_\textrm{PCR}-1}\pt{2N'_I+1}\pt{N'_A+1}+G_\textrm{PCR}N_{V_\textrm{PCR}}+\mu_1^2/2,\label{paramgausspcr}
    \end{split}
\end{equation}
resulting in the expression given for $R_\textrm{PCR}$ in the main text.

%

\end{document}